\documentclass[11pt]{article}

\usepackage[boxed]{algorithm2e}
\dontprintsemicolon

\usepackage{multirow}
\usepackage{bigstrut}
\usepackage{xspace}
\usepackage{comment}

\usepackage{cite}

\newcommand{\alert}[1]{\typeout{ALERT(\thepage): #1}}

\newcommand{\newloglike}[2]{\newcommand{#1}{\mathop{\rm #2}\nolimits}}
\newloglike{\E}{E}

\newcommand{\etal}{{\it et al{.}\/}\xspace}
\newcommand{\buzz}[1]{\textbf{#1}}

\newcommand{\suppress}[1]{}

\newcommand{\ratifymsg}{\mbox{\textit{ratify}}}

\newcommand{\newop}[2]{\newcommand{#1}{\mbox{\textsc{#2}}}}
\newop{\sendop}{send}
\newop{\receiveop}{receive}
\newop{\readop}{read}
\newop{\writeop}{write}
\newop{\flipop}{coin-flip}

\SetKwFunction{CoinFlip}{CoinFlip}

\title{Randomized Protocols for Asynchronous Consensus}
\author{James Aspnes\thanks{
Department of Computer Science, Yale University,
New Haven, CT 06520-8285, USA.
Email: {\tt aspnes-james@cs.yale.edu}.
Supported by NSF grants CCR-9820888 and CCR-0098078.}}

\begin{document}

\maketitle

\begin{abstract}
The famous Fischer, Lynch, and Paterson impossibility proof 
shows that it is impossible to solve the consensus problem
in a natural model of an asynchronous distributed system if even a
single process can fail.
Since its publication, 
two decades of work on fault-tolerant asynchronous consensus algorithms 
have
evaded this impossibility
result by using extended models that provide 
(a) randomization,
(b) additional timing assumptions,
(c) failure detectors, or
(d) stronger synchronization mechanisms than are available in the
basic model.
Concentrating on the first of these approaches, 
we illustrate the history and structure of randomized
asynchronous consensus protocols by giving detailed descriptions of several
such protocols.
\end{abstract}


\section{Introduction}
\label{section-introduction}

The \buzz{consensus} problem~\cite{PeaseSL1980}
is to get a group of $n$ processes in a
distributed system to agree on a value.  
A \buzz{consensus protocol}
is an algorithm that produces such an agreement.
Each process in a consensus protocol has, as part of its initial
state,
an input from some specified range,
and must eventually \buzz{decide} on some output from the same range.
Deciding on an output is irrevocable; though a process 
that has decided may continue to
participate in the protocol, it cannot change its decision value.
The restricted problem in which the input range is $\{0,1\}$ is called
\buzz{binary consensus}.  Except as otherwise noted, all of the
protocols discussed hereafter are binary consensus protocols.

Correct consensus protocols must
satisfy the following three conditions:
\begin{enumerate}
\item \buzz{Agreement}.  All processes that decide choose the same value.
\item \buzz{Termination}.  All non-faulty processes eventually decide.
\item \buzz{Validity}.  The common output value is an input value of
some process.
\end{enumerate}

This is not precisely the definition originally given by
Pease, Shostak, and Lamport~\cite{PeaseSL1980}. 
Their paper used the even stronger
condition of \buzz{interactive consistency}, in which
all non-faulty
processes compute the same vector of purported inputs,
and this vector correctly identifies the input of each non-faulty
process.  But the definition above is the one that is generally
accepted today.
It derives from the three-part definition used by
Fischer, Lynch, and Paterson~\cite{FischerLP1985}, 
though in their paper
the validity condition is replaced by a much weaker
\buzz{non-triviality} condition.  Non-triviality 
says only that there exist
different executions of the protocol that produce different common
decision values.  Non-triviality is implied by validity (consider an
execution in which all processes have input $0$ versus one in which all
process have input $1$), but it is less useful for applications, since
it says nothing about the relationship between inputs and outputs.
Nonetheless, non-triviality is enough to show that consensus is
impossible in the usual model.

The \buzz{Fischer-Lynch-Paterson} (\buzz{FLP})
impossibility result~\cite{FischerLP1985} demonstrates that there is no
deterministic
protocol that satisfies the agreement, termination, and non-triviality
conditions for an asynchronous message-passing system in which any
single process can fail undetectably.  
A similar result, proved by Loui and Abu-Amara~\cite{LouiA1987} using
essentially the
same technique, shows that consensus is also impossible in an
asynchronous shared-memory system with at least one undetectable 
failure.
More details of these and similar results, and of the models in which
they apply, can be found in the survey by Fich and Ruppert appearing
elsewhere in this volume.\alert{should have better cite for Fich-Ruppert}

And yet we would like to be able to solve consensus.
To escape from the FLP result, we must extend the model in some way.
Several such extensions
that have been used to solve asynchronous
consensus are described in
Section~\ref{section-extensions}.
In this paper, we concentrate on the use of randomized algorithms,
and give a more detailed description of
randomized approaches
in 
Sections~\ref{section-randomization} through~\ref{section-wait-free}.

\section{Extensions to the model}
\label{section-extensions}

The extensions to the base model that have been used to
circumvent the FLP result can be divided into four classes.  
\begin{enumerate}
\item\label{extension-randomization}
\buzz{Randomization}.
Randomized models provide probabilities for some transitions.  This
means that instead of looking at a single worst-case execution, one
must consider a probability distribution over bad executions.
If the termination requirement is weakened to require termination only
with probability $1$, the FLP argument no longer forbids consensus:
non-terminating executions continue to exist, but they may collectively
occur only with probability $0$.

There are two ways that randomness can be brought into the model.  One
is to assume that the model itself is randomized; instead of allowing
arbitrary applicable operations to occur in each state, particular operations
only occur with some probability.  
Such a \buzz{randomized scheduling} approach was first proposed by
Bracha and Toueg
\cite{BrachaT1985} (who called their version \buzz{fair scheduling}).
A recent attempt to revive this approach can be found
in the \buzz{noisy scheduling} model of~\cite{Aspnes2000}.
Randomized scheduling allows for very simple algorithms;
unfortunately, it depends on assumptions about the behavior of the
world that may not be justified in practice.
Thus it has not been as popular as the second approach, in which
randomness is located in the processes themselves.

In this
\buzz{randomized algorithm} approach,
processes are equipped with coin-flip operations that return random
values according to some specified probability distribution.
Again, we can no longer talk about a single worst-case execution, but
must define a probability distribution on executions.
Defining this distribution requires nailing down all the other
nondeterminism in the system (i.e., the order in which messages are
sent and delivered, or in which shared-memory operations are
performed), which is done formally by specifying an \buzz{adversary}.
An adversary is a function from partial executions 
to operations that says which applicable operation to carry out at
each step.
(Details are given in Section~\ref{section-randomized-model}.)
Given an adversary, the result of the coin-flip operations are
the only remaining unknowns in determining which
execution occurs.  So we can assign to each set of executions the
probability of the set of sequences of
coin-flip outcomes that generate them.

The adversary we have just described is the \buzz{strong adversary};
it can observe the entire history of the system, including past
coin-flip outcomes and the states of processes and the communications
mechanism, but it cannot predict future coin-flip outcomes.
A strong adversary gives a weak model in which consensus is possible
but difficult.
Weakening the adversary gives a stronger model (in the sense of
granting more strength to the processes); many consensus protocols
have been designed for \buzz{weak adversaries} 
with a restricted view of the system.

This survey concentrates primarily on randomized algorithms,
largely because they lie closest to the domain of expertise of the
author, but also because they require the least support from the
underlying system.
Protocols using randomization are discussed starting in
Section~\ref{section-randomization}.
An excellent survey on work using randomization up to 1989 can be
found in~\cite{ChorD1989}.

\item\label{extension-timing-assumptions}
\buzz{Timing assumptions.}
Consensus can be achieved despite the FLP result by adding timing
assumptions to the model that exclude bad executions.
Dolev, Dwork, and Stockmeyer~\cite{DolevDS1987} characterize the effects of
adding limited synchrony assumptions to a message-passing system, and show
in particular
that consensus becomes possible with up to $n$ faulty processes under
a variety of restrictions on the order in which processes take steps
or messages are delivered.
Dwork, Lynch, and Stockmeyer~\cite{DworkLS1988} introduced the
\buzz{partial synchrony} model, in which either there is a bound on message
delay that is not known to the processes, or there is a known
bound that applies only after some initial time $T_0$.  They
describe consensus protocols for this model that work with a bound on
the number of faulty processes.
Attiya, Dwork, Lynch, and Stockmeyer~\cite{AttiyaDLS1994} give a still
more refined model in which there are known bounds $C$ on the ratio
between the maximum and minimum real time between steps of the same process
and $d$ on the maximum message delay; under these assumptions, they
prove an upper bound of $(f+1)Cd$ and a lower bound of $(f+1)d$ on the
amount of real time needed to solve consensus with $f$ faulty
processes.  Their upper bound uses timeouts to detect failures and can
be seen as an early example of the failure detector approach described
below.

In the shared-memory framework,
a model in which processes can deliberately delay operations
in order to avoid overwriting each other's values was
used to obtain a very simple and efficient consensus protocol by Alur,
Attiya, and Taubenfeld~\cite{AlurAT1997}.
More recently, Anderson and Moir~\cite{AndersonM1999} have used
scheduling assumptions from the operating systems world to design
protocols that run in constant time when processes run on a
uniprocessor under a priority-based scheduling regime or under one that
provides minimum scheduling quanta.

\item\label{extension-failure-detectors}
\buzz{Failure detectors}.
With failure detectors, some mechanism exists for notifying
other processes that a process has failed.
An example of a failure detector is the timeout-based mechanism used
by Dwork, Lynch, and Stockmeyer~\cite{DworkLS1988}.
Much more interesting are \buzz{unreliable failure detectors}, where
the failure detector can misidentify faulty processes as non-faulty
and vice versa.
Work on protocols for unreliable failure detectors was initiated by
Chandra and Toueg~\cite{ChandraT1996}.
Chandra, Hadzilacos, and Toueg~\cite{ChandraHT1996} extended this work
by showing the minimum conditions an unreliable failure detector must
satisfy to permit consensus.
Being able to detect failures, even unreliably, 
permits solving consensus by
electing a coordinator, without the danger of having the
protocol hang forever waiting for a failed co-ordinator to wake up. 
Further examples of work on failure detectors and their limitations in various
models can be found in~\cite{DolevFKM1997,AguileraCT2000}.

\item\label{extension-strong-objects}
\buzz{Strong primitives}.
In these models,
stronger shared-memory
primitives extend or supplement
the basic 
operations
of
reading and writing registers.
Loui and Abu-Amara~\cite{LouiA1987} showed that consensus is solvable
with one (but not two) failures using test-and-set bits and is
solvable for arbitrarily many failures using three-valued
read-modify-write registers.  Extending this work,
Herlihy~\cite{Herlihy1991} defined a hierarchy of shared-memory
objects based on their \emph{consensus number}, defined somewhat
informally as the maximum number of processes for which the object
can solve wait-free consensus.  The essence of this line of research 
is that
consensus is used as a test
problem to prove that certain shared-memory objects cannot implement
other, stronger objects.  
More robust
definitions have appeared in subsequent work
(see, for example, \cite{Jayanti1997}), and there is now the beginning
of a broad theory of the power of shared-memory
objects (e.g., \cite{Ruppert1997,Ruppert1998,Ruppert1999,AfekMT1999}).

Herlihy's paper also gave one of the first \emph{universal
constructions} for arbitrary shared-memory objects based on consensus;
showing that objects that can solve consensus with arbitrarily many
failures can implement arbitrary shared objects.
Another construction is due to
Plotkin~\cite{Plotkin1989}, based on \emph{sticky bits} similar to the
3-valued read-modify-write registers of Loui and
Abu-Amara~\cite{LouiA1987}.
Subsequent constructions have shown
how to implement consensus or protocols of equivalent power
using primitives such as
load-linked/store-conditional~\cite{Herlihy1993,ShavitT1995}.
\end{enumerate}

Some consensus protocols combine aspects of protocols from different
models.  For example, a hybrid protocol of Aguilera and
Toueg~\cite{AguileraT1998} solves consensus very quickly given a
failure detector, but solves it eventually using randomization 
if the failure detector doesn't work.

\section{Consensus using randomization}
\label{section-randomization}

One of the first approaches to solving consensus despite the FLP
result was to use randomization.  
The goal of a randomized consensus protocol is to give the
set of 
non-terminating executions a probability of zero.  This does not, in a
sense, require breaking the FLP result: these zero-probability
executions continue to exist.  However, they are irrelevant in
practice, provided one is willing to accept a probabilistic
termination guarantee.
This requires modifying the model as described in
Section~\ref{section-randomized-model}.

The first randomized consensus protocol was given by
Ben-Or~\cite{Ben-Or1983}; as it provides much of the structure for
later protocols, we give a description of it in
Section~\ref{section-ben-or}.  Later papers
extended Ben-Or's work in two directions: the literature on
message-passing consensus protocols has largely concentrated on
solving agreement problems using cryptographic techniques or private
channels with a linear bound on the number of faulty processes
(including processes with Byzantine faults, which may misbehave
arbitrarily instead of simply stopping); while work in shared-memory
systems has used the underlying reliability of the shared-memory
system to solve consensus in the wait-free case, where there no limit
on how many processes may fail 
but failures are limited to crash failures.  
We describe some of the Byzantine agreement work briefly in
Section~\ref{section-ben-or-successors} and discuss wait-free
shared-memory algorithms at greater length in
Section~\ref{section-wait-free}.

\section{How randomization affects the model}
\label{section-randomized-model}

Adding randomization involves changing both the model, to include the
effect of random inputs, and the termination condition, to permit
non-terminating executions provided all such executions together have
probability zero.\footnote{Permitting probability-zero non-terminating
executions is not required in a synchronous
model, where algorithms exist that use randomization to obtain high
efficiency but that still 
terminate after finitely many rounds in all
executions~\cite{GoldreichP1990,Zamsky1996}.
The difference can be accounted for by the fact that deterministic
fault-tolerant consensus is possible in a synchronous model.
In an asynchronous model, if a randomized consensus protocol
terminated in all executions, we could simply replace all coin-flips
with hard-wired constants and get a deterministic consensus protocol,
contradicting the FLP result.}

From the point of view of the processes, the main change in the model
is the addition of a new $\flipop$ operation.
The $\flipop$ operation behaves a bit like a $\readop$
operation, except that it returns a random value to the process that
executes it
instead of the value from some register.  
Depending on the precise details of the model, this value might be the
outcome of a fair coin-flip, or might be a value chosen from a larger
range.

Adding $\flipop$ operations requires additional changes to the model
to handle probabilities of executions.
In each state of the system there may be
a large number of operations of different processes that may occur
next.
Previously the choice of which of these operations 
occurs
has
been implicit in the choice of a single execution, 
but now we want to consider ensembles consisting of many executions,
where the probabilities of individual executions
are determined by the return values of the
$\flipop$ operations.
In this framework,
it is convenient to assign control of which operation
occurs in each configuration to an explicit \buzz{adversary}, a
function from partial executions to operations.
The idea is that the adversary always chooses what operation happens next,
but if that operation is a $\flipop$, the result of the operation is
random.

The adversary thus takes on the role of the single worst-case
execution in the deterministic model.  It also takes on the
responsibilities of the worst-case execution, in that it must guarantee
the fairness conditions required by the algorithm.
Once the adversary is specified, which execution occurs is determined
completely by the outcome of $\flipop$ operations, and the probability
of an execution or set of executions is determined by the probability
of the corresponding coin-flips.
We can then talk about the \buzz{worst-case expected complexity} of an
algorithm, 
by considering the worst possible adversary, and then
taking the average of some
complexity
measure
over all executions weighted by their probability given this
adversary.\footnote{Technically, it may be that in some
models no single worst-case adversary exists; or in other words, that 
for any
adversary, there is a slightly nastier adversary that produces worse
performance.  An example might be when running under a fairness
condition that requires an adversary to eventually permit some
operation, but allows different adversaries to delay the operation for 
arbitrarily long finite times at increasing cost to the algorithm.
To handle such possibilities, one can define the
worst-case expected complexity instead as the supremum over all adversaries 
of the expected complexity.}

Defining the adversary as a function from partial executions to
applicable operations means that the adversary can in effect see the
entire history of the execution, including outcomes of past
coin-flips, internal states of the processes, and the contents of
messages and memory locations.
Restricting the adversary's knowledge
provides an opening for further variations on the model.

The adversary defined above is called 
the \buzz{strong adversary}.
The strong adversary has the advantage of mathematical simplicity, but
for practical purposes it may be \emph{too} strong.
While 
the choice of which operation occurs next in a system might
depend on the configuration in a complicated way (for example, whether or not
reading a particular register causes a page fault might be very
difficult to predict without examining the entire previous execution),
one can reasonably argue that Nature is not so malicious that only the
strong adversary can encompass its awful power.  This observation has
motivated the development of a variety of \buzz{weak adversaries} that
permit faster consensus protocols.\footnote{Conventionally, weak
adversaries give rise to strong models, and strong adversaries to weak
models; the strength or weakness of the model is a function of how
much power it gives to the processes.}  A weak adversary
might be unable to break cryptographic tools used by the processes, or
it might be restricted more directly by not being allowed to observe
anything (such as message or register contents) that would not affect the
scheduling mechanism in a real system.
Formally, this usually involves restricting the adversary based on a
notion of \buzz{equivalent executions}.
Two executions are
equivalent if they consist of the same sequence of operations
(ignoring parameters
of operations and their return values), and the adversary must choose
the same operation (again ignoring parameters) after any two equivalent
partial executions.
An adversary that is restricted in this way
is called a \buzz{content-oblivious} adversary,
since it cannot see the contents of processes, messages, or registers.
A still weaker \buzz{oblivious} adversary cannot even distinguish
between different operations; it chooses in each state only which
process takes the next step.
Other models restrict the actions of the adversary, by imposing
additional timing assumptions or perturbing the adversary's choices
randomly.
We describe some of
these weak-adversary models in Section~\ref{section-weak}.

In addition to changing the model, we also adapt the correctness
conditions for a consensus protocol to reflect the presence of
randomness.  In particular, the universal quantifier in the
termination requirement, which previously spanned all executions, now
instead
spans all adversaries; and we only require for any particular
adversary that the protocol terminate with probability 1.

A similar change in quantification would not be useful
for the agreement and validity requirements.  Since any violation of
agreement or validity must occur after finitely many steps---and thus
finitely many coin-flips---any such violation would occur with
nonzero probability.
Thus we continue to
demand that 
the agreement and validity conditions
hold in all executions, which is equivalent to
demanding that they hold with probability $1$.\footnote{It does change
the problem to allow disagreement with nonzero probability.  A version
of the Byzantine agreement problem
that permits disagreement
but seeks to minimize its probability
is studied by
Graham and Yao~\cite{GrahamY1989}.  They cite an earlier unpublished paper of
Karlin and Yao as originating this approach.}

\section{Fault-tolerant message-passing protocols}
\label{section-message-passing}

This section describes fault-tolerant message-passing protocols.
These achieve consensus given a bound on the number of faults (which
may vary from protocol to protocol).  We begin with Ben-Or's
original exponential-time protocol in Section~\ref{section-ben-or} and
describe some of its faster descendants in
Section~\ref{section-ben-or-successors}.

\subsection{Ben-Or's protocol}
\label{section-ben-or}

Ben-Or's protocol is the earliest protocol that achieves consensus
with probabilistic termination in a model with a strong adversary.
Designed for a message-passing system, it tolerates $t < n/2$ crash
failures, and requires
exponential expected time to converge in the worst case.

Each process starts off with a \buzz{preference} equal to its input.
The protocol proceeds in rounds, each of which has two stages, a
voting stage and a ratification stage.  Nothing in the system
guarantees that all processes proceed through different rounds at the
same time; instead, each process keeps track of its own round and uses
round labels on messages from other processes to decide whether to
use them in its current round, throw them away (for messages 
from rounds the process has already finished), or save them for later
(for messages from rounds the process has not yet reached).

The first stage of each round implements a voting procedure; each
process transmits its current preference $p$ to all processes
(including itself) by sending a message of the form $(1,r,p)$, and
then waits to receive $n-t$ such messages.  If any process receives
more than $n/2$ votes for a single value, it causes all processes to
decide on this value using the ratification mechanism.

This mechanism is implemented by the second stage of each round.  Any
process that has observed a majority of votes for $v$ sends a message
$(2,r,v,\ratifymsg)$ to all processes.  A process that has not
observed a majority for either value sends instead a place-holder
message $(2,r,?)$.

As in the first stage, each process waits to receive at least $n-t$
second-stage messages.
Any process that receives even a single $(2,r,v,\ratifymsg)$ message
in round $r$ changes its preference for round $r+1$ to $v$.  If, in
addition, it receives more than $t$ such messages, it immediately
decides on $v$.  If, on the other hand, it
receives only $(2,r,?)$ messages, it flips a fair coin to choose a new
preference for the next round.  The process then continues with the
first stage of round $r+1$.

This procedure is summarized as Algorithm~\ref{alg-ben-or}.
\begin{algorithm}[tp]
\caption{Ben-Or's consensus protocol.  Adapted from \cite{Ben-Or1983}.}
\label{alg-ben-or}
\SetKwFunction{BenOrConsensus}{BenOrConsensus}
\KwIn{boolean value input}
\KwOut{boolean value stored in output}
\KwData{boolean preference, integer round}
\Begin{
    preference $\leftarrow$ input\;
    round $\leftarrow$ 1\;
    \While{true}{
        send (1, round, preference) to all processes\;
        wait to receive $n-t$ (1, round, $\ast$) messages\;
        \eIf{received more than $n/2$ (1, round, v) messages}{
            send (2, round, v, $\ratifymsg$) to all processes
        }{
            send (2, round, ?) to all processes
        }
        wait to receive $n-t$ (2, round, $\ast$) messages\;
        \eIf{received a (2, round, v, $\ratifymsg$) message}{
            preference $\leftarrow$ v\;
            \If{received more than $t$ (2, round, v, $\ratifymsg$) messages}{
                output $\leftarrow$ v
            }
        }{
            preference $\leftarrow$ \CoinFlip{}
        }
        round $\leftarrow$ round + 1\;
    }
}
\end{algorithm}

The algorithm guarantees agreement because:
\begin{enumerate}
\item At most one value can receive a majority of votes in the first
stage of a round, so for any two messages $(2,r,v,\ratifymsg)$ 
and $(2,r,v',\ratifymsg)$, $v=v'$.
\item If some process sees $t+1$ $(2,r,v,\ratifymsg)$ messages, then
every process sees at least one $(2,r,v,\ratifymsg)$ message.
\item If every process sees a $(2,r,v,\ratifymsg)$ message, every
process votes for $v$ in the first stage of round $r+1$ and every
process that has not
already decided decides $v$ in round $r+1$.
\end{enumerate}

Validity follows by a similar argument; if all processes vote for the
their common input $v$ in round $1$, then all processes send
$(2,r,v,\ratifymsg)$ and decide in the second stage of round $1$.

Assuming a weak adversary that cannot observe the contents of
messages,
termination follows because if no process decides in round $r$, then
each process either chooses its new preference based on the majority
value $v$ in a $(2,r,v,\ratifymsg)$ message; or it chooses its new
preference randomly, and there is a nonzero probability that all of
these random choices equal the unique first-stage majority value (or
each other, if there is no majority value).  
The situation is more complicated with a strong adversary, as
different processes may be in the first and second stages at the same
time, and so the
first-stage majority value may not be determined until after the
adversary has seen some of the coin-flips from the second stage.
The algorithm continues to work with a strong adversary, but a much
more sophisticated proof is needed~\cite{AguileraT1998TR}.

Unfortunately, in either case the probability that the algorithm
terminates in any given round 
may be exponentially small as a function of the number of processes, 
requiring exponentially many rounds.
Note also that each process continues to run the protocol even
after deciding; however, the protocol can
be modified so that each process exits at most one round after first setting
its output value.

\subsection{Faster protocols}
\label{section-ben-or-successors}

Ben-Or's protocol not only showed that consensus becomes possible with
randomization, but also initiated a large body of work on randomized
protocols for the harder problem of \buzz{Byzantine agreement}, in
which faulty processes can exhibit arbitrary behavior instead of
simply crashing.  
This work has generally assumed the availability of cryptographic
tools.
Rabin~\cite{Rabin1983} showed that Byzantine
agreement can be solved in constant expected time given a shared
coin visible to all processes, and described an implementation of such
a coin based on digital signatures and a trusted dealer.
Feldman and Micali~\cite{FeldmanM1997} 
gave a constant-round shared-coin for a synchronous system that
uses secret sharing to avoid the need for a trusted dealer.

A constant-time shared coin for an \emph{asynchronous}
system was given by Canetti and Rabin~\cite{CanettiR1993} based in part on
further unpublished work by Feldman.
For the Canetti and Rabin protocol the cryptographic assumptions can
be replaced by the assumption of private channels.  A simplified
presentation of the Canetti-Rabin protocol
that tolerates crash failures only is given
in~\cite[Section 14.3.2]{AttiyaW1998}.

\section{Wait-free shared-memory protocols}
\label{section-wait-free}

A protocol that tolerates up to $n-1$ crash
failures is called \buzz{wait-free}, because it means that any process
can finish the protocol in isolation without waiting for the others.
Wait-free message-passing protocols for all but trivial problems are
easily shown to be impossible, as a process running in isolation
cannot tell whether it is the sole survivor or simply the victim of a
network partition~\cite{BrachaT1985}.

With shared memory, things become easier.  
Even though processes may fail, it is usually assumed that data in
the shared memory
survives.\footnote{See~\cite{AfekGMT1995,JayantiCT1998} for
models in which the shared memory itself can be faulty.}
This eliminates the possibility of partition--- even a process running
in isolation can still read the notes left behind by dead processes
that ran before.  The consensus problem then becomes a problem of
getting the contents of the shared memory into a state that
unambiguously determines the decision value, even for processes that
may have slept through most of the protocol.

In this section, we begin describing the history of wait-free
shared-memory consensus, and then give examples of different
approaches to the problem, showing the range of the trade-off between
efficiency and robustness against adversaries of increasing strength.

The first randomized protocol to use shared memory to solve consensus 
was described by Chor, Israeli, and
Li~\cite{ChorIL1994}, for a model in which processes can generate
random values and write them to shared memory in a single atomic
operation (this is equivalent to assuming a weak adversary that cannot
see
internal coin-flips until they are written out).  
A noteworthy feature of the protocol is that it works for
values in any input range, not just binary values.

We describe the Chor-Israeli-Li protocol in more detail in
Section~\ref{section-CIL}.  To give a brief summary, the essential
idea
of the Chor-Israeli-Li protocol is to have the processes run a race,
where processes advance through a sequence of rounds as in the Ben-Or
protocol, and slow processes adopt the preferences of faster processes
that have already reached later rounds.  If a single process
eventually pulls far
enough ahead of all processes that disagree with it, 
both this \buzz{leader} and the other processes can
safely decide on the leader's preference, knowing that any other
processes will adopt the same preference by the time they catch up.
Processes flip coins to decide whether or not to advance to the next
level.  With the probability of advancement set at $\frac{1}{2n}$, a
leader emerges after $O(n^2)$ total work.

With a strong adversary, the Chor-Israeli-Li protocol fails.  The winning
strategy for the adversary is to construct a ``lockstep execution''
that keeps all processes at the same round, by stopping any process
that has incremented its round until all other processes have done so
as well.  This strategy necessarily requires that the adversary be
able to observe internal states of the processes when making its
scheduling decisions.  It is still possible to solve consensus with a
strong adversary, but a different approach is
needed.

Wait-free consensus protocols that tolerate a strong adversary began
to appear soon after the publication of the conference version of the
Chor-Israeli-Li paper.
Abrahamson~\cite{Abrahamson1988} gave the first randomized wait-free
consensus protocol for a strong adversary.  Though described in terms
of processes acquiring locks of increasing strengths, if the strengths
of the locks are interpreted as round numbers it superficially resembles
the Ben-Or protocol translated to shared memory.  As in Ben-Or's
protocol, the method for obtaining agreement is to have the processes
choose new preferences at random; after $2^{O(n^2)}$ steps on average,
the processes' random choices will have agreed for a long enough
period that the adversary cannot manipulate them into further
disagreement.

Also as with Ben-Or's protocol, eliminating exponential waiting
time required replacing the independent local coins of the processes
with global coin protocols shared between the processes. 
The first protocol to do this was given by Aspnes and
Herlihy~\cite{AspnesH90}; it used a round structure similar to the
Chor-Israeli-Li protocol but relied on a shared coin protocol based on
majority voting to shake the processes into agreement.
The Aspnes and Herlihy protocol was still fairly expensive, requiring
an expected $O(n^4)$ total operations in the worst case, though
much of this cost was accounted for by the overhead of using a
primitive $O(n^2)$-work
snapshot subroutine constructed specifically for the protocol.  
Subsequent work with the strong adversary
by many 
authors~\cite{AttiyaDS1989,SaksSW91,Aspnes93,DworkHPW1999,BrachaR1990,BrachaR1991,AspnesW1996}
has largely retained the overall structure of the Aspnes and Herlihy
protocol, while reducing the overhead and eliminating in some cases
annoyances like the use of unboundedly large registers.
The net result of these developments was to reduce the expected total work to
achieve consensus to $O(n^2 \log n)$ using the shared-coin protocol of
Bracha and Rachman~\cite{BrachaR1991} and the
expected per-process work to $O(n \log^2 n)$ using the shared-coin
protocol of~\cite{AspnesW1996}.
An $O(n^2 \log n)$ consensus protocol using the 
Bracha-Rachman coin is described in
Section~\ref{section-shared-coins}.

Further improvements in the strong-adversary model stalled at this
point.  It was later shown by Aspnes~\cite{Aspnes1998} that no
strong-adversary consensus protocol could run in less than
$\Omega(n^2/\log^2 n)$ total work in essentially any model in which
the FLP bound applies.
The essential idea of this lower bound was to extend the
classification of states in the FLP argument 
as bivalent or univalent to a randomized framework, by defining a
state as bivalent if the adversary can force either decision value to
occur with high
probability, univalent if it can force only one decision value to
occur with
high probability, and \buzz{null-valent} if it can force
neither decision value to occur with high probability.
This last case is equivalent to saying that the adversary cannot bias
the outcome of the protocol too much--- or, in other words, that the
protocol acts like a shared coin.
Aspnes showed, using a variant of the FLP argument, that any consensus
protocol that reaches a decision in less than $n^2$ steps is likely to 
pass
through a null-valent state,
and provided a separate lower bound on
shared coins to show that $\Omega(n^2/\log^2 n)$ expected work
(specifically, $\Omega(n^2/\log^2 n)$ expected write operations)
would be needed to
finish any protocol starting in a null-valent state.
Combining these two results gives the lower bound on consensus.

Even before this lower bound was known,
the lack of further improvement in strong-adversary protocols led to
greater interest in protocols for weak adversaries.
Aumann and Bender~\cite{AumannB1996} gave a shared coin algorithm for
the
\buzz{value-oblivious adversary} 
that cannot observe the internal states of the processes or values
that have been written to memory but not yet read.  Based on 
propagating values through a butterfly network, their algorithm gives
a constant-bias shared coin in $O(n \log^2 n)$ total work.
Concurrently,
Chandra devised an algorithm for repeatedly solving consensus in a
model with essentially the same adversary, with a
polylogarithmic per-process time bound.
Chandra's protocol uses a stockpile of pre-flipped
coins that the processes agree to use.  The initial execution of the
protocol is expensive, due to the need to generate an initial
stockpile of unused coins, but subsequent executions can solve new
instances of the consensus problem in only $O(\log^2 n)$
time.  Chandra's algorithm also gives a very streamlined
implementation of the rounds mechanism from earlier strong-adversary
protocols, reducing the shared data needed to just two arrays of
multi-writer bits.  (We use a version of this algorithm to reduce
consensus to shared coin in
Section~\ref{section-consensus-from-shared-coin}.)
Soon afterwards, Aumann~\cite{Aumann1997} showed how to achieve 
$O(\log n)$ expected per-process work even for a single iteration of
consensus.
It is not clear
whether $O(\log n)$ expected steps is the best possible in this model,
or whether further improvements may be obtained.

\subsection{Weak-adversary protocols}
\label{section-weak}

In this section, we first describe the Chor-Israeli-Li protocol
that demonstrated the possibility of wait-free consensus, and then
sketch out some more recent work that uses a similar approach.  The
unifying theme of these protocols is to have the processes run a race
where advancement to the next phase is controlled by some random
process and winning the race (by getting far enough ahead of the other
processes) determines the outcome of the protocol.  Although the
adversary can use its control over timing to handicap particular
processes, a weak adversary cannot identify which phase each process
is in and thus cannot prevent a victor from emerging.

\subsubsection{The Chor-Israeli-Li protocol}
\label{section-CIL}

Pseudocode for a simplified version of
the Chor-Israeli-Li protocol is given as Algorithm~\ref{alg-cil}.
\begin{algorithm}[tp]
\caption{Chor-Israeli-Li protocol.  Adapted with modifications
from \cite{ChorIL1994}.}
\label{alg-cil}
\SetKwIf{WithProb}{WithProbElse}{with probability}{do}{else}{end}
\SetKwInput{LocalData}{Local data}
\SetKwInput{SharedData}{Shared data}
\KwIn{input (an arbitrary value)}
\KwOut{return value}
\LocalData{preference, round, maxround}
\SharedData{one single-writer multi-reader register for each process}
\Begin{
    preference $\leftarrow$ input\;
    round $\leftarrow$ 1\;
    \While{true}{
        \lnl{line-CIL-write}        
        write (preference, round)\;
        \lnl{line-CIL-collect}
        read all registers $R$\;
        maxround $\leftarrow$ $\max_{R} R.\mbox{round}$\;
        \lnl{line-CIL-win-test}
        \eIf{for all $R$ where $R.\mbox{round} \ge \mbox{maxround} - 1$,
            $R.\mbox{preference} = \mbox{v}$}{
            \KwRet{v}
        }{
            \lnl{line-CIL-follower}
            \If{exists $v$ such that
                    for all $R$ where $R.\mbox{round} = \mbox{maxround}$,
                    $R.\mbox{preference} = \mbox{v}$}{
                \lnl{line-CIL-follower-end}
                preference $\leftarrow v$\;
            }
            \lnl{line-CIL-advance}
            \WithProb{$\frac{1}{2n}$}{
                round $\leftarrow$ $\max(\mbox{round}+1, \mbox{maxround}-2)$
            }
        }
    }
}
\end{algorithm}

Communication between processes is done by having each process
alternate between writing out its current round and preference to its
own output
register in Line~\ref{line-CIL-write}, 
and reading all the other processes' registers to
observe their recent states in Line~\ref{line-CIL-collect}. 
The only interactions with the shared memory are in these two lines.
A process notices that the race has been won
if it observes that processes with preference $v$ are far enough
ahead of all disagreeing processes in
Line~\ref{line-CIL-win-test}; in this
case, it decides on $v$ and exits.
If the process does not decide, it adopts the common preference of the
fastest processes provided they all agree
(Lines~\ref{line-CIL-follower}--\ref{line-CIL-follower-end}).

The only tricky part is ensuring that eventually some process does win
the race, i.e.{} moves far enough ahead of any processes that disagree
with it.  
This is done in
Line~\ref{line-CIL-advance}
by having each process choose whether to
advance to the next round at random.
Chor~\etal show that,
provided the adversary cannot delay a process's write depending on its
choice to advance or not, a leader
emerges on average after $O(n)$ passes through the loop.
The expected total work is $O(n^2)$, since each pass requires $O(n)$
read operations.

\subsubsection{Protocols for still weaker adversaries}

For still weaker adversaries, it is possible to remove the
randomization from the Chor-Israeli-Li protocol and still solve consensus.
This is essentially what is done in the uniprocessor
consensus protocol of Anderson and Moir~\cite{AndersonM1999}, which
relies on quantum and/or priority-based scheduling to avoid lockstep
executions and which achieves consensus deterministically in \emph{constant}
work
per process; and in the ``noisy environment'' consensus
protocol of Aspnes~\cite{Aspnes2000}, which assumes that the
adversary's schedule is perturbed by cumulative random noise and
achieves consensus in expected $O(\log n)$ work per process.  

\subsection{Strong-adversary protocols}
\label{section-shared-coins}

The main tool for defeating a strong adversary---one that can react
to the internal states of processes and the contents of memory---has
been the use of \buzz{weak shared coin} protocols.
These provide a weak shared coin abstraction
with the
property that, regardless of the adversary's behavior,
for each value $0$ or $1$
there is a constant minimum probability $\epsilon$
that all processes agree on that value as the value of the shared coin.
The coin is said to be weak because $\epsilon$ is in general less than
$\frac{1}{2}$: there will be some executions in which the adversary
either seizes control of the coin or prevents agreement altogether.

We describe a typical weak shared coin protocol in
Section~\ref{section-bracha-rachman}.  As in most such protocols, the
shared coin is obtained by taking a majority of many local
coin flips generated by individual processes.
While the adversary can bias the outcome of the coin by selectively
killing processes that are planning to vote the wrong way, it can only
hide up to $n-1$ votes in this way, and with enough votes it is likely
that the majority value will not shift as a result of the adversary's
interference.

Before presenting a shared coin protocol, we will show that having
such a protocol does indeed give a solution to consensus.

\subsubsection{Consensus from a shared coin}
\label{section-consensus-from-shared-coin}

Given a polynomial-work shared coin protocol, it is easy to build a
wait-free shared-memory consensus protocol requiring similar total
work.  
The basic idea is the same as in the Ben-Or protocol: disagreements
are eliminated by sending all but the fastest processes off to flip
coins.
The actual structure of the algorithm
resembles the Chor-Israeli-Li algorithm, 
in that processes proceed through a sequence of rounds,
and slow processes adopt the preferences of faster ones,
but now 
the rounds structure is no longer used to distinguish winners
from losers but instead simply ensures that the fastest processes do
not have to wait for processes stuck in earlier
rounds.
With appropriate machinery, we can arrange that in each round
all processes either (a) think they are leaders and agree with all
other leaders, or (b) participate in the shared coin protocol.  Since
all processes in the first category agree with each other, and all
processes in the second category will also choose this agreed-upon
value with probability at least $\epsilon$, after $O(1/\epsilon) =
O(1)$ rounds all processes agree.

The first wait-free consensus protocol to use this technique was the
protocol of Aspnes and Herlihy~\cite{AspnesH90}, which included an
ad-hoc snapshot algorithm and various other additional mechanisms to
allow the protocol to be built from single-writer registers using the
techniques of the day.  
A more recent algorithm due to
Chandra~\cite{Chandra1996} gives a
much simpler implementation of the 
multi-round framework using two arrays of multi-writer
bits.

Pseudocode for a simplified version of this algorithm is given as
Algorithm~\ref{alg-chandra-ladder}.  The 
main simplification the use of unbounded
bit-vectors, which avoids some additional machinery in Chandra's algorithm for
truncating the protocol if it has not terminated in $O(\log n)$ rounds 
and switching to a slower bounded-space algorithm.\footnote{This
trick, of switching from a fast algorithm that is running too long
to a slower but bounded algorithm
was originally devised by Goldreich and
Petrank~\cite{GoldreichP1990} for synchronous Byzantine agreement protocols.
Chandra's original algorithm
was designed for a weak adversary and its shared coin subroutine
consumes ``pre-flipped''
shared coins stored in
memory.  
Switching to a second algorithm is needed to avoid running out of
these pre-flipped coins.
But since the switch occurs
with low probability, the cost of the slower algorithm is rarely
incurred, and thus does not change the total asymptotic expected
cost.}
The simplified algorithm requires very few assumptions about the
system.  In particular, it works even when processes do not have
identities~\cite{BuhrmanPSV2000} and with infinitely many
processes~\cite{AspnesSS2002}.
\begin{algorithm}[tp]
\caption{Wait-free consensus using shared coins.  Adapted with
modifications from \cite{Chandra1996}, Figure 1.}
\label{alg-chandra-ladder}
\SetKwFunction{Consensus}{Consensus}
\SetKwFunction{SharedCoin}{SharedCoin}
\SetKwInput{LocalData}{Local data}
\SetKwInput{SharedData}{Shared data}
\SetKwInput{Subroutines}{Subprotocols}
\KwIn{boolean value input}
\LocalData{boolean preference $p$, integer round $r$, boolean new
preference $p'$}
\SharedData{boolean mark$[b][i]$ for each $b \in \{0,1\}$, $i \in
Z^+$, 
of which mark$[0][0]$ and mark$[1][0]$ are initialized to true
while
all other elements are initialized to false.}
\Subroutines{Shared coin protocols
$\SharedCoin_r$ for $r = 0, 1, \ldots$.}
\Begin{
    $p \leftarrow$ input\;
    $r \leftarrow$ 1\;
    \While{true}{
        \lnl{line-Chandra-write}
        mark$[p][r] \leftarrow$ true\;

        \uIf{mark$[1-p][r+1]$}{
            \lnl{line-Chandra-behind}
            $p' \leftarrow 1-p$\;
        }
        \lElse{
            \uIf{mark$[1-p][r]$}{
                \lnl{line-Chandra-tied}
                $p' \leftarrow \SharedCoin_r()$\;
            }
            \lElse{
                \uIf{mark$[1-p][r-1]$}{
                    \lnl{line-Chandra-ahead}
                    $p' \leftarrow p$\;
                }
                \Else{
                    \lnl{line-Chandra-decide}
                    \KwRet $p$\;
                }
            }
        }

        \If{mark$[p][r+1] = $ false}{
            \lnl{line-Chandra-new-preference}
            $p \leftarrow p'$\;
        }
        \lnl{line-Chandra-advance}
        $r \leftarrow r+1$\;
    }
}
\end{algorithm}

We will not give a detailed proof of correctness of
Algorithm~\ref{alg-chandra-ladder}, referring the interested reader
instead to~\cite{Chandra1996}.  However, we can give some intuition
about why it works.  

The two arrays of bits substitute for the round fields in the
Chor-Israeli-Li algorithm, allowing a process to quickly determine if
its preference
is in the lead without having to read $n$ separate registers.
A process $P$ that reaches round $r$ with preference $p$ 
registers this fact by setting mark$[p][r]$ in
Line~\ref{line-Chandra-write}.

What $P$ does next depends on whether it sees itself behind, tied with,
one round ahead, or two rounds ahead of the fastest process with the
opposite preference:
\begin{itemize}
\item If it sees a mark in mark$[1-p][r+1]$, it knows that it is behind, and
switches to the other preference (Line~\ref{line-Chandra-behind}).
\item If the latest mark it sees is in mark$[1-p][r]$, it assumes it is
tied, and chooses a new preference using the shared coin
(Line~\ref{line-Chandra-tied}).
\item If the latest mark it sees is in mark$[1-p][r-1]$, 
it keeps its current preference (Line~\ref{line-Chandra-ahead})
but does not yet decide, as it may be that some process $Q$ with the
opposite preference
is close enough that $Q$ will immediately set mark$[1-p][r]$ and
then think it is tied with $P$.
\item Finally, if it sees no mark later than mark$[1-p][r-2]$ (which it doesn't
bother to read), then any $Q$ with a different preference has not yet
executed Line~\ref{line-Chandra-write} in round $r-1$; and so after
$Q$ has done so, it will see the mark that $P$ already put in
mark$[p][r]$, and then switch its preference to $P$'s preference $p$.  
In this case $P$ can safely decide $p$
(Line~\ref{line-Chandra-decide}) knowing that any slower process
will switch to the same preference by the time it catches up.
\end{itemize}
As a last step, the process checks to see if at least one process with
its old preference has already advanced to the next round.  In this
case, it discards its decisions from the current round and sticks with
its previous preference; otherwise, it adopts the new preference it
determined in the preceding lines (Line~\ref{line-Chandra-new-preference}).  
This avoids a problem where some process $P$ decides
$p$ in $r$, but a process $P'$ with the same preference
later sees a tie in $r-1$ 
(after some other processes catch up with it),
executes the shared coin, and
suddenly switches its preference while catching up with $P$.

Intuitively, agreement holds precisely because of the explanations for
Lines~\ref{line-Chandra-decide} and~\ref{line-Chandra-new-preference}.
But a full correctness proof is subtle,
and depends among other things on the precise order in which
mark$[1-p][r+1]$, mark$[1-p][r]$, and mark$[1-p][r-1]$ are read in
Algorithm~\ref{alg-chandra-ladder}.
However, if one accepts that the algorithm satisfies agreement, it is
trivial to see that it also satisfies validity, because no process
every changes its preference unless it first sees that some other
process has a different preferences.

Termination follows from the properties of the shared coin.
Note that it is possible
that some processes tied for the lead will skip the coin
because they read mark$[1-p][r]$ just before the others write it.
But it is not hard to show that these processes will at least agree
with each other.
So the processes that do participate in the coin will
fall into agreement with
the non-participants with at least a constant probability, due to the
bounded bias of the coin, 
and agreement is reached after a constant number of rounds on
average.  Since the overhead of the consensus protocol is small (five
operations per round in the worst case), the cost is dominated by the
cost of the shared coin protocol.

\subsubsection{Bracha and Rachman's shared coin protocol}
\label{section-bracha-rachman}

We can use any shared coin subroutine we like in Chandra's protocol; as
discussed previously, the expected cost of the algorithm will be
within a constant factor of the cost of the shared coin, provided the
shared coin guarantees at most constant bias.
The most efficient shared coin protocol
currently known for
the strong adversary, when measured according to expected total work,
is Bracha and Rachman's 1991 
voting protocol~\cite{BrachaR1991}.  Pseudocode for
this protocol is given as Algorithm~\ref{alg-bracha-rachman}.
\begin{algorithm}[tp]
\caption{Bracha and Rachman's voting protocol.  Adapted from
\cite{BrachaR1991}, Figure 2.}
\label{alg-bracha-rachman}
\SetKwInput{LocalData}{Local data}
\SetKwInput{SharedData}{Shared data}
\KwIn{none}
\KwOut{boolean output}
\LocalData{boolean preference $p$; integer round $r$; utility
variables $c$, total, and ones}
\SharedData{single-writer register
$r[p]$ for each process $p$, each of which holds a pair of integers
(flips, ones),
initially $(0,0)$}
\Begin{
    \Repeat{total $> n^2$}{
        \For{$i \leftarrow 1$ \KwTo $\frac{n}{\log n}$}{
            \lnl{line-br-coin-flip}
            $c \leftarrow \CoinFlip()$\;
            \lnl{line-br-write}
            $r[p] \leftarrow (r[p].\mbox{flips}+1, r[p].\mbox{ones}+c)$\;
        }
        \lnl{line-br-termination-read}
        Read all registers $r[p]$\;
        total $\leftarrow \sum_p r[p].\mbox{flips}$\;
    }
    \lnl{line-br-last-read}
    Read all registers $r[p]$\;
    total $\leftarrow \sum_p r[p].\mbox{flips}$\;
    ones $\leftarrow \sum_p r[p].\mbox{ones}$\;
    \lnl{line-br-test}
    \eIf{$\frac{\mbox{total}}{\mbox{ones}} \ge \frac{1}{2}$}{
        \KwRet{$1$}\;
    }{
        \KwRet{$0$}\;
    }
}
\end{algorithm}

The intuition behind this protocol is the same as for all voting-based
protocols: The processes collectively generate many ``common votes,''
which in this case consist of all votes generated before 
$n^2$ votes have been written to the registers.  Each process's view
of the common votes is obscured both by additional ``extra votes''
that are generated by processes that have not yet noticed that there
are enough votes in the registers, and by the adversary's selective
removal of ``hidden votes'' by delaying processes between generating
votes in Line~\ref{line-br-coin-flip} and writing them out in
Line~\ref{line-br-write}.\footnote{Our explanation of the
Bracha-Rachman protocol follows the
analysis of a similar protocol
from~\cite{AspnesW1996}.  Bracha and Rachman's original analysis 
in~\cite{BrachaR1991} uses a slightly different classification that
includes common and extra votes but does not separate out the issue of
hidden votes.  Their classification requires the analysis of a more
sophisticated random process than the one considered here.}
The reason the protocol works is that we
can argue that the $n^2$ common votes have at least a constant probability
of giving a majority large enough that neither the random drift of up
to $n^2/\log n$ extra votes nor the selective pressure of up to $n-1$
hidden votes is likely to change the apparent outcome.

Because the extra votes are not biased by the adversary, they are less
dangerous than the hidden votes and we can tolerate more of
them.  This is why the protocol can amortize the cost of the $n$ read
operations to detect termination in
Line~\ref{line-br-termination-read} over the $\frac{n}{\log n}$ votes
generated in the inside loop.  This amortized termination test was the
main contribution of the Bracha and Rachman protocol, and was what
finally brought the expected total work for consensus down to nearly 
$O(n^2)$ from the $O(n^3)$ (or worse) bounds on previous protocols.

In detail,
the votes are classified as follows.
These classes are not exclusive; a vote that is in either of the first
two classes may also be in the last.
\begin{enumerate}
\item \emph{Common votes} consist of all votes generated before the
sum of the $r[p]$.total fields exceeds $n^2$.
In a sense,
these are all the votes that would be seen by all processes if they had been
written out
immediately.  There will be between $n^2+1$ and $n^2+n$ such votes.
\item \emph{Extra votes for process $P$} are those votes $X_i$ that
are not part of the common votes and that are generated by some
process $Q$ before $P$ reads $r[Q]$
in
Line~\ref{line-br-last-read}.  
Each process $Q$ contributes at most $\frac{n}{\log n}$ such extra
votes, because it cannot generate more without executing the
termination test staring in Line~\ref{line-br-termination-read}.
The common votes plus the extra votes
for $P$ include all votes that $P$ would have seen had they been
written out
immediately.
\item \emph{Hidden votes for $P$} are those votes which were generated
by some process $Q$ but not written to $r[Q]$ when $P$ reads $r[Q]$.
Each process $Q$ contributes at most one hidden vote for $P$.
\end{enumerate}

The total vote for $P$ is given by:
\begin{displaymath}
\mbox{(common votes)}
+
\mbox{(extra votes for $P$)}
-
\mbox{(hidden votes for $P$)}.
\end{displaymath}

When the adversary permits a process to flip its coin in
Line~\ref{line-br-coin-flip}, it is already determined whether or not
that coin-flip will count towards the common votes or the extra votes
for any particular process $P$.  
So both the common votes and the extra votes consist of a sequence of
unbiased fair coins, and the only power the adversary has over them is
the choice of when to stop the sequence.

Using the normal approximation to the binomial distribution, it is
possible to show that the net majority for $1$ of the approximately $n^2$
common votes is at least $3n$
with some constant probability $p$.
Adding in the $\frac{n^2}{\log n}$ extra votes for a particular
process $P$ may adjust this total
up or down; it reduces it below $n$ only if at some point during an
unbiased
random walk of length $\frac{n^2}{\log n}$ the total drops below $-2n$.
Standard results\footnote{E.g.~\cite[Theorem III.7.3]{Feller1970}.}
show that this probability is bounded by
$\frac{1}{n^2}$, so even when we multiply the probability for a
single process by the number of processes $n$, the probability that the 
extra votes are less than $n$ for \emph{any} $P$ is still less than
$\frac{1}{n}$.  We thus have a probability of at least 
$p\left(1-\frac{1}{n}\right)$ that the common votes plus the extra
votes for all $P$ are at least $n$.
Since each $P$ has at most $n-1$ hidden votes, each $P$ then sees a
positive net vote and decides $1$.

The preceding argument shows that when $n > 1$, all processes decide
$1$ with at least a constant probability $\epsilon = p/2$.  The 
case for decision value $0$ is symmetric.

The total work of Algorithm~\ref{alg-bracha-rachman} is $O(n^2 \log
n)$; there are $O(n^2)$ votes cast by all of the processes together,
and each vote has an amortized cost of $O(\log n)$.
Plugging the Bracha-Rachman shared coin into
Algorithm~\ref{alg-chandra-ladder} thus gives a consensus protocol
whose expected total work is also $O(n^2 \log n)$.

This is the current best known bound on expected
total work for wait-free consensus in the 
strong-adversary model.
Since the Bracha-Rachman algorithm, the only further improvement in this
model has been the Aspnes-Waarts shared coin~\cite{AspnesW1996}, which modifies
the Bracha-Rachman coin to prevent any single process from having to perform
more than $O(n \log^2 n)$ work, at the cost of increasing the total
work bound to $O(n^2 \log^2 n)$.  

There is still some room left to improve
the total work, but not much.  We have previously mentioned the
$\Omega(n^2/\log^2
n)$ lower bound on the expected number of write operations for any
wait-free consensus protocol 
from~\cite{Aspnes1998}.  
The same paper conjectured that the actual lower
bound is $\Omega(n^2/\log n)$.
Using this conjectured lower bound, 
and throwing in an extra logarithmic factor for
the cost of amortizing reads over coin-flips and writes,
a reasonable guess for the true cost of consensus in this model
might be $\Theta(n^2)$.

\section{Acknowledgments}

The author would like to thank Prasad Jayanti and Sam Toueg for
many useful comments on the original abstract and outline of this
survey, the students in Yale's Fall 2001 Theory of Distributed
Systems class for their patience with an early draft of some of the
later sections, Hagit Attiya and Sergio Rajsbaum for suggesting some 
references, 
the anonymous referees for their extensive and useful comments,
and Marcos Aguilera for identifying and providing fixes
for several serious technical errors.

\bibliographystyle{plain}
\bibliography{paper}
\end{document}